# CHEMICAL SYNTHESIS AND MATERIALS DISCOVERY

Anthony K. Cheetham,[1,2] Ram Seshadri[1] and Fred Wudl[1]

[1] *Materials Research Laboratory, University of California; Santa Barbara, California 93106, USA.*
[2] *Department of Materials Science and Engineering*
*National University of Singapore; 9 Engineering Drive 1, 117575 Singapore.*

*Corresponding author: akc30@cam.ac.uk

*Abstract*

Functional materials impact every area of our lives ranging from electronic and computing devices to transportation and health. In this Perspective, we examine the relationship between synthetic discoveries and the scientific breakthroughs that they have enabled. By tracing the development of some important examples, we explore how and why the materials were initially synthesized and how their utility was subsequently recognised. Three common pathways to materials breakthroughs are identified. In a small number of cases, such as the aluminosilicate zeolite catalyst ZSM-5, an important advance is made by using design principles based upon earlier work. There are also rare cases of breakthroughs that are serendipitous, such as the buckyball and Teflon[R]. Most commonly, however, the breakthrough repurposes a compound that is already known and was often made out of curiosity or for a different application. Typically, the synthetic discovery precedes the discovery of functionality by many decades; key examples include conducting polymers, topological insulators and electrodes for lithium-ion batteries.

*Introduction*

Our Perspective explores the various ways in which the synthesis of novel compositions of matter with unique crystalline structures can lead to important advances in materials discovery along with technological applications. The initial synthesis itself is a chemical discovery, and it usually precedes the realisation of its associated functionality, which could be considered a discovery in the materials science area. In a few cases, the latter arises when a well-known compound is obtained with a new morphology, often at the nanoscale. The formation of graphene from graphite, and the creation of quantum dots from semiconductors such as CdSe, are important examples. We shall show that, in many cases, the breakthrough in materials science takes place in an area that is unrelated to the motivations for the original synthesis. For example, lithium cobalt oxide, $Li_xCoO_2$, which has evolved since the 1980s into a major family of battery cathodes for rechargeable lithium batteries, was originally studied in the 1950s for its unusual magnetic properties. Indeed, we find many examples in which the time-lag between the original chemical synthesis and the realisation of important materials applications is several decades.

Our examples include unsung heroes, such as aluminosilicate zeolite catalysts, which play crucial roles in the chemical and petrochemical industries, among others, but have never attracted major recognition in terms of awards such as Nobel prizes. By contrast, high-profile discoveries such as the high temperature cuprate superconductors and the Buckyball, $C_{60}$, have been honoured with Nobel prizes but have not yet had the technological impact that was initially anticipated. This is not very surprising because important scientific discoveries do not necessarily lead directly to commercial applications. Nevertheless, in their different ways, all of these different types of breakthroughs have a major influence on science and the scientific community. **Figure 1** illustrates some diverse examples of chemical discoveries, along with the schematic to show the technological area in which they have subsequently had an impact. Each of these cases will be discussed in more detail below.

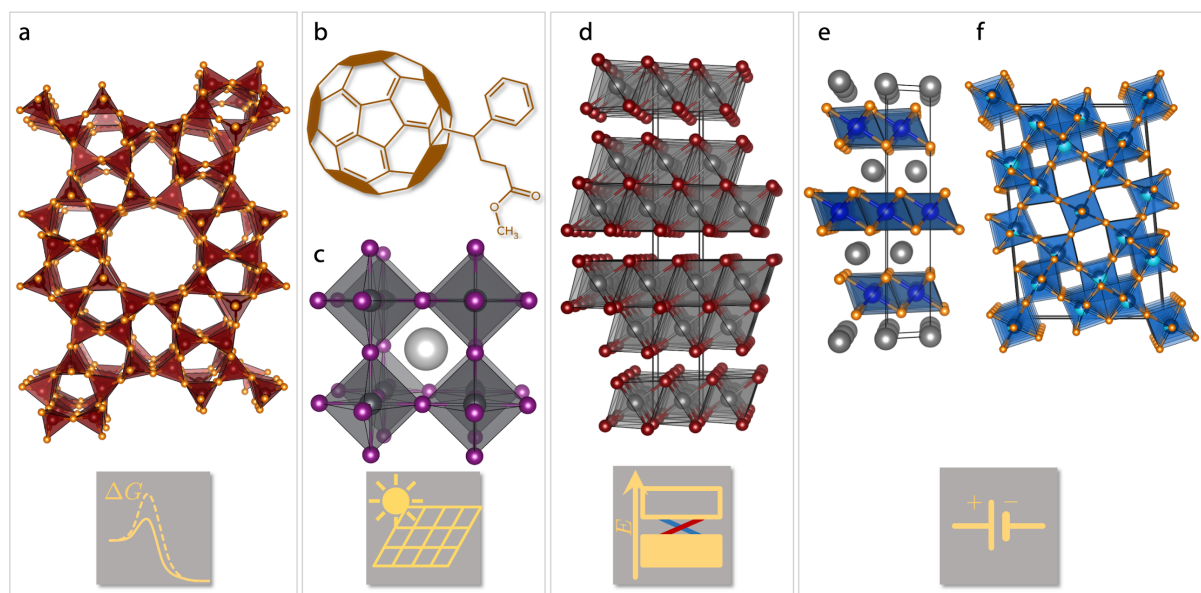

**Figure 1:** A depiction of some of the functional materials discussed in this perspective, along with schematics to illustrate their main fields of application. (a) The crystal structure of the zeolite catalyst ZSM-5. (b) The molecular structure of [6,6]-phenyl-$C_{61}$-butyric acid methyl ester (PCBM), a fullerene compound that is the workhorse n-doped materials in bulk heterojunction polymer photovoltaics. (c) Idealised structure of the halide perovskite semiconductor (MA)PbI$_3$ (MA = methylammonium, $CH_3NH_3$) (d) The crystal structure of $Bi_2Te_3$, a topological insulator. (e) The crystal structure of $Li_xCoO_2$, a key cathode material for lithium-ion batteries (LIBs). (f) The crystal structure of the shear-phase compound, $TiNb_2O_7$, an emerging anode material for fast-charging LIBs.

We have identified three common pathways to technological breakthroughs based on the synthesis of novel compositions of matter. At one end of the spectrum, there are materials that were made as a consequence of design principles established from earlier successes and a well-defined material need. A significant fraction of materials discovery is conducted in this way, following the mandates of funding bodies that prescribe specific missions and goals. However, we find that very few major breakthroughs are made in this manner, though incremental advances can lead to important practical applications. At the other end of the spectrum are compounds that were discovered serendipitously by scientists who were looking for something entirely different, but who had the vision to realise the implications of their unexpected findings. In between, there are compounds that have sat on the shelf for many years until their functionality and commercial potential were finally recognised. Some examples of each pathway are discussed in the sections below and the timelines of some important cases are illustrated in **Figure 2**. We also explore developments in data mining and artificial intelligence for the enabling of materials discovery and discuss whether high-throughput computational approaches, coupled with high-throughput synthesis and AI/ML methods, will eventually replace the role of chemical intuition and exploratory synthesis.

*Materials discoveries using design principles*

In practice, novel and exciting compositions of matter are rarely found by using design principles developed earlier, though large numbers of incremental advances – some very significant – have been made in this way. Aluminosilicate zeolites provide some interesting examples, including the eye-catching case of zeolite ZSM-5, an entirely new polymorph of $A_x Si_{1-x}Al_xO_2$ ($A$ is typically a univalent metal cation or a proton). ZSM-5 was synthesised using an amine-template methodology that was widely adopted in the zeolite community. By using a rather large quaternary amine cation, tetrapropylammonium, a group at Mobil made not only a new zeolite architecture [**Figure 1(a)**], but a material with a much higher Si/Al ratio than typical zeolites.[1] This higher Si/Al ratio led to enhanced thermal and chemical stability, as a consequence of which ZSM-5 has found widespread commercial use as an acid catalyst for high temperature processes such as xylene isomerisation and hydrocarbon cracking. Another landmark discovery in the porous materials area was the synthesis of the first nanoporous polymorphs of $AlPO_4$, which exploited the known relationship between the structures of α-quartz and berlinite, $AlPO_4$, while again adopting the amine template strategy.[2] This templating strategy is still used today to create zeolitic architectures and chemistries, such as the germanosilicate ITQ-56.[3] In a closely-related, game-changing discovery, self-assembling surfactant amine templates were used successfully in the early 1990s to make mesoporous solids, especially silicas,[4] leading to an exciting new class of functional materials.

In the world of organic materials, tetrathiafulvalene (TTF) was designed specifically for its electrical properties, such as redox behaviour and conductivity.[5] TTF was the precursor to the first organic metal[6] and its derivatives eventually produced the first of many organic superconductors. Over the years, the TTF backbone has been incorporated into many organic molecules that utilize its easily accessible redox behaviour.[7] From the area of polymers, another example is polyethylenedioxythiophene (PEDOT). PEDOT was designed and synthesized at Bayer AG specifically for several of its thin film conductivity and antistatic properties.[8,9] The latter is its major bulk application.[9] More recently it was found to form thermoelectric films with figures of merit ($ZT$) as high as 0.4.[10] By far its most important applications, however, are in organic electronics as hole transporting layers and as transparent thin film conductors with conductivities as high as 4600 S/cm,[11] far surpassing those of indium tin oxide (ITO).

*Serendipitous discoveries*

The discovery of important new materials by serendipity is extremely rare, but the examples of Teflon and the Buckyball, $C_{60}$, show that it does occasionally happen. The discovery of poly(tetrafluoroethylene) (TEFLON$^R$) was acknowledged to be serendipitous by its discoverer, Roy Plunkett.[12] A group of chemists and chemical engineers at the DuPont Company were working on chlorofluorocarbon refrigerant gases in the 1930s. These gases were mostly based on single carbon atom molecules, but Plunkett was working on more complex 2-carbon atom molecules derived from tetrafluoroethylene (TFE). He prepared a large amount of TFE and stored it in a metal gas cylinder to be converted at a later date to a chlorofluorocarbon. Months later, when Plunkett needed some TFE for the synthesis of these gases, there was no gas pressure in the cylinder, yet it did not lose any weight! To see what had happened, he and his co-worker cut the cylinder open and discovered that the inside contained a white, waxy solid that was insoluble in all common solvents and inert to most chemicals. Plunkett concluded that the gas had polymerized completely on the metal surface of the cylinder. Though there are refrigerants based on chlorotrifluoroethylene, there appear to be no commercial refrigerants based on TFE.[13] The discovery of the buckyball, $C_{60}$, followed a series of earlier ideas and observations that, with the knowledge of hindsight, might have led to its characterization before the famous 1985 experiment that resulted in the 1996 Nobel Prize in Chemistry for Curl, Kroto and Smalley. For example, the concept of hollow carbon molecules had been suggested by David Jones (a.k.a. Daedalus) in the New Scientist in 1966,[14] and there were theoretical studies that predicted a stable, aromatic cage of 60 carbon atoms in 1970.[15] In addition, work by Kaldor and his colleagues at Exxon in 1984 on carbon soots that build up on hydrocarbon cracking catalysts revealed a large number of clusters with even numbers of carbon atoms between approximately $C_{40}$ and $C_{100}$.[16] $C_{60}$ did not attract attention as being special, though its mass spectrometry signature was about 20% stronger than the neighbouring clusters.[16] The pre-eminence of $C_{60}$ among the many clusters that had been observed by Kaldor was first recognized by Curl, Kroto, Smalley and their co-workers in a search for medium sized carbon molecules that might be found in interstellar space.[17] The experimental conditions in their graphite vapourisation chamber, prior to detection by mass spectrometry, enabled them to obtain a more quantitative distribution of carbon clusters in which $C_{60}$ (and to a lesser extent $C_{70}$) was dominant. The subsequent development of a synthesis that yielded a bulk solid product[18] enabled a wide range of exciting developments, such as the preparation of superconducting fullerides.[19] Even more importantly, the discovery of $C_{60}$ arguably led to the nanotechnology revolution and the discoveries of other nanoscale carbons, particularly carbon nanotubes and graphene. In an interesting recent twist to this story, it has now been confirmed spectroscopically that $C_{60}$ is actually present in space where it contributes to the fine structure of the Diffuse Interstellar Bands.[20]

*Delayed impact of known compounds*

The majority of breakthroughs in materials science involve the application of compounds that were already known and had either been initially synthesized out of curiosity or were repurposed for a new application. We have assembled a selection of important examples, both inorganic and organic, with their timelines in (**Figure 2**) and we discuss four of these cases in more detail below.

*Polyacetylene*: Although it was initially observed as a highly crosslinked black powder called *Cuprene* by Marcellin Berthelot in 1866,[21] polyacetylene ($CH_x$) was first synthesized by a known polymer synthetic method by Giulio Natta in the 1950s.[22] He obtained a pure, linear crystalline form that was good enough for preliminary structure determination by powder X-ray diffraction. Unfortunately, its powder morphology did not facilitate determination of its physical properties, especially electron transport. The powder morphology also precluded any possible technological applications, though pressed pellets of the crystalline powder exhibited a resistivity of $10^{10}\,\Omega\,\text{cm}$, which is considerably lower than a typical insulating polyolefin ($10^{18}\,\Omega\,\text{cm}$).

In the mid 1970's, Alan MacDiarmid described his work with Alan Heeger on $(SN)_x$ during a visit to Tokyo and showed his beautiful films to the audience. After the talk, Hideki Shirakawa showed him a silvery film that consisted only of carbon and hydrogen. MacDiarmid was so impressed that he invited Shirakawa to visit the University of Pennsylvania and, as a consequence, in 1977 they published the revolutionary result of the dramatic increase in conductivity upon doping $CH_x$ with iodine.[23] The crucial finding that $CH_x$ could be prepared in thin film form had already been reported by Ito, Shirakawa and Ikeda in 1974.[24] However, it was also known, thanks to Berthelot and Natta, that polyacetylene was a very unstable material, so it was clear that $CH_x$ was going to have limited technological applications. On account of this, the publication of the 1977 paper triggered an immediate examination of other conjugated polymers *vis-a-vis* their doping-based electrical conductivity. Thus, the whole field of electrooptical applications of conjugated polymers was born, including the 1990 discovery of the first polymer LEDs that were based on the insoluble and intractable PPV.[25] Shortly thereafter, the soluble and processable poly[2-methoxy-5-(2-ethylhexyloxy)-1,4-phenylenevinylene] (MEH-PPV) was designed and synthesized[26] With the contemporaneous discovery of $C_{60}$ as an electron acceptor, it was clear that a this electron poor molecule could be "doped" into MEH-PPV to produce a semiconductor polymer that resulted in an unusual organic photovoltaic. However, $C_{60}$ is only sparingly soluble in organic solvents, so a solution to this problem was found with the discovery of [6.6] - phenyl $C_{61}$ butyric acid methyl ester [PCBM, **Figure 1(b)**], a derivative that was more soluble. Its doping into MEH-PPV produced a much higher photovoltaic efficiency and the discovery of the bulk heterojunction[27] phenomenon that is ubiquitous in organic photovoltaics. Conducting polymers continue to find applications in diverse areas, most recently in thermoelectrics[28] and wearables.[29]

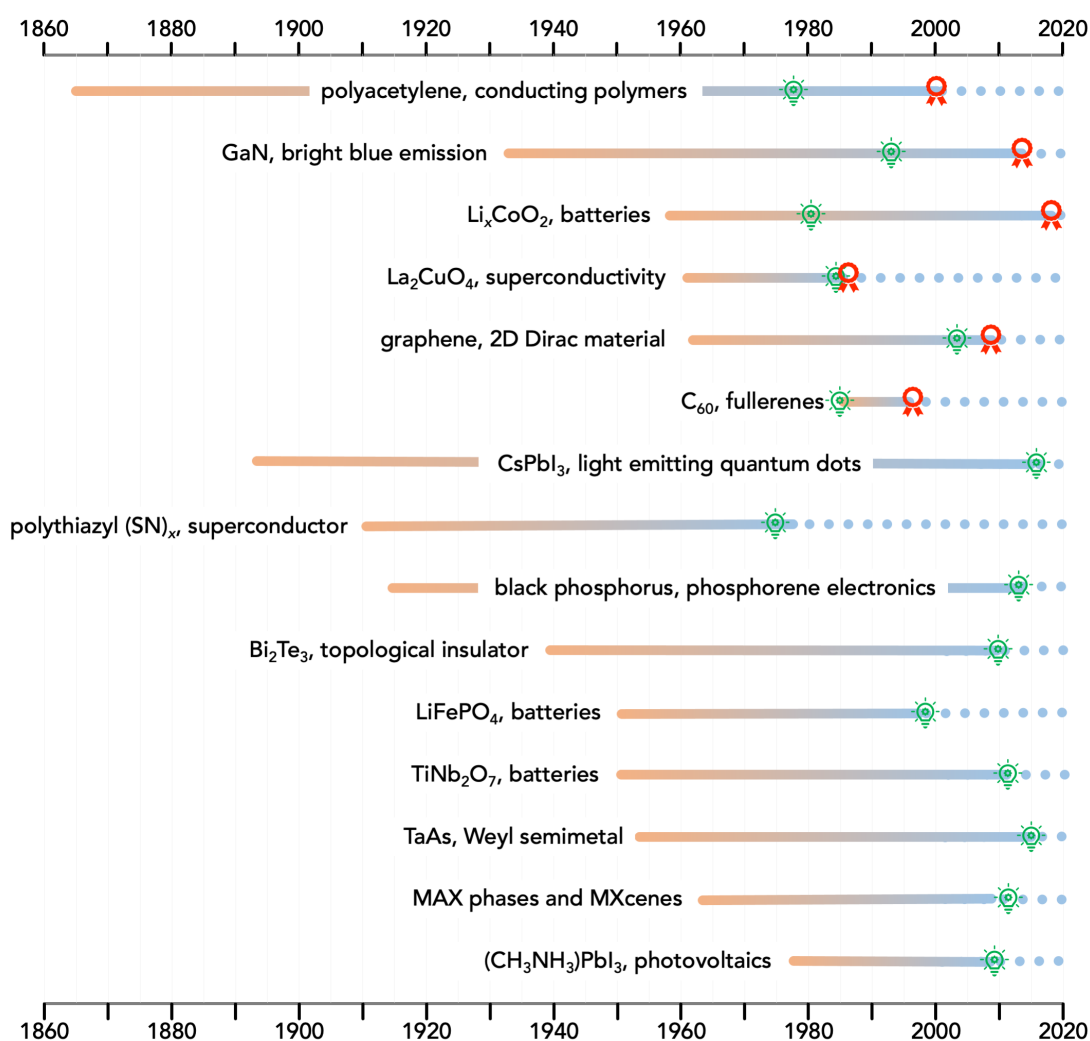

**Figure 2:** Timelines for some important functional materials with delayed impact. The left-hand extreme of each line (in brown) shows the year of the original synthesis, while the year in which the key discovery of functionality was published is shown by a green bulb. For the cases that resulted in a Nobel Prize, the year of the prize is indicated by a red laurel. The on-going work following the key discovery is shown by the blue dots. Such on-going work is often incremental, but in some cases, it results in significant optimisations and commercial applications.

***(MA)PbX$_3$ (MA = methylammonium, CH$_3$NH$_3$; X = Cl, Br, I):*** Hybrid 3D perovskites of general formula (MA)PbX$_3$ [**Figure 1(c)**] were first synthesized out of curiosity by Weber at the University of Stuttgart in 1978.[30] Weber noted that while the chloride was colourless, the bromide and iodide were intensely coloured, though he also claimed that they did not exhibit significant electronic conductivity. Furthermore, it was reported that the structures were cubic. Initially, these new compounds did not attract very much attention, but subsequent work in 1985 by solid state nuclear magnetic resonance[31] revealed a series of temperature dependent phase transitions associated with dynamical disorder of the methylammonium cations. Shortly afterwards, it was found that the phase transitions in (MA)Pb$X_3$ involved a lowering of symmetry on cooling from cubic to tetragonal and finally othorhombic forms.[32] Importantly, it was also found that the bromide and iodide compounds were semiconducting[33,34] and that the conductivity was higher for the tin analogues of the lead compounds. Mitzi and coworkers at IBM were notably the next to advance the properties of the hybrid tin halides, which were shown to display dimensionality-controlled insulator-to-metal transitions[35] and potential applications in thin-film electronics.[36]

The alkylammonium lead halides were subsequently thrust into prominence in 2009, 31 years after their initial discovery, when the group of Miyasaki at the University of Tokyo revealed that $CH_3NH_3PbBr_3$ and $CH_3NH_3PbI_3$ could be used as visible-light sensitizers in photoelectrochemical cells, with the iodide showing a solar conversion efficiency of 3.8 %.[37] This exciting development captured the attention of many other groups around the world, and within only 2-3 years efficiencies exceeding 10 % were being reported. A huge amount of work aimed at increasing the efficiency ensued during the 2010s, with optimisation around the device architecture as well as the use of mixed cations (for example, caesium or formamidinium (FA) can be used as alternatives to MA) and mixed anions, usually Br with I. The performance with single junction cells is now around 25 %,[38] and that figure approaches 30% when the halide perovskite is combined with silicon in a tandem cell configuration.[39]

There remain many challenges in the deployment of lead halide perovskites for photovoltaic cells, including long term chemical stability issues and environmental concerns around the use of lead. Tin performs well, but the chemical stability of Sn(II) is more challenging than that of Pb(II). Lead-free hybrid double perovskites such as $(MA)_2AgBiBr_6$, have recently been made,[40] but it has proven difficult to make the iodide analogues with the required lower band gaps; it is easy to satisfy the Tolerance Factor criterion, but the $M^{III}$ cations are usually too small to meet the size criterion for octahedral coordination with iodides.[41] Nevertheless, the field of hybrid halide perovskites is still thriving 44 years after the initial discovery of (MA)PbI$_3$ and many more exciting developments are anticipated.

*Topological Quantum Materials:* Materials have long been understood on the basis of the symmetries they display. Transitions between different material phases — a paramagnet to a ferromagnet, or a metal to a superconductor as examples — are understood in terms of certain symmetries being broken. For example, the continuous symmetry of spins in the paramagnetic state, associated with their being able to point in any direction of the crystal, is broken when the crystal becomes ferromagnetic upon cooling and the spins are obliged to take in specific (lower symmetry) arrangements. The 2016 Nobel Prize in Physics, awarded to Thouless, Haldane, and Kosterlitz, recognized that — beyond symmetry — topology can play a role in our understanding of materials and their properties. Several features of materials, for example, the manner in which electronic energy bands cross one-another, or the manner in which collections of spins assemble into complex textures, can be associated with topological indices. In topological quantum materials, the indices calculated on the inside of the crystal may not correspond with the indices calculated for a vacuum. This sudden change of the topological index at the interface between the crystal and the vacuum can give rise to surface states associated with the crystal being "cut" by analogy with the need to cut a donut (topological index: the genus = 1, for the single hole) in order to transform it into a doughball (genus = 0 for the absence of a hole).[42]

The prediction of the existence of materials with so-called $Z_2$ topological order[40,41] led to the realization that certain valence-precise semiconductors with high-atomic number elements (a prerequisite for strong spin-orbit coupling) could display the characteristics of a topological insulator, with a bulk semiconducting gap but with spin-polarized conducting states at the surface. Some obvious materials choices, with the added benefit of being layered and cleaving readily (important for surface science studies), are the $Bi^{3+}$ chalcogenides such as $Bi_2Te_3$, the structure of which is displayed in **Figure 1(d)**. The structure of $Bi_2Te_3$, today a well-known thermoelectric material, has been known since 1939[45] and the existence of Bi:Te compounds in the ratio 2:3 since at least 1905.[46] An examination of the band structure of these compounds[47] confirmed the existence of the requisite surface states in insulating $Bi_2Te_3$, establishing it to be a topological insulator.[48]

Following hot on the heels of topological insulators came the prediction and realization of Weyl semimetals: crystals that host Weyl fermions are characterized by arcs connecting Weyl points on the Fermi surface. Weyl fermions have long been known as massless quasiparticles that could arise from solutions to the relativistic Dirac equation, but their experimental realization has been elusive. The compound TaAs, which crystallizes in the $I4_1md$ space group and lacks inversion symmetry, was structurally characterized in 1963.[49] An examination of its electronic structure indeed suggested the features expected for a Weyl semimetal,[50] and this has since been experimentally verified.[51,52]

***Lithium Cobalt Oxide:*** Lithium cobalt oxide, $Li_xCo_{1-x}O$, was first prepared in the 1950s[53] at the Westinghouse Research Laboratories in Pittsburg PA at a time when there was general interest in layered structures of the transition metals of general formula $Li_xM_{1-x}O_2$, $M$ = Mn, Ni, Cu. The primary focus of the $Li_xCo_{1-x}O$ study, aside from establishing the composition range and crystal structure, was to understand the role played by changing the ratios of $Co^{2+}$ and $Co^{3+}$ in determining the magnetic properties. It was found that the structure was cubic in the range 0 < $x$ < 0.2, like CoO, but rhombohedral and layered at $x$ = 0.5, i.e. $LiCoO_2$ [**Figure 1(e)**]. More than 20 years later and shortly after the first demonstration of using a reversible lithium insertion compound, $Li_xTiS_2$, as the cathode in a rechargeable battery,[54] Goodenough and co-workers at Oxford selected $LiCoO_2$ as the cathode for their oxide-based lithium ion battery (LIB).[55] The choice of oxides was made on the basis that the oxide would give a higher cell voltage and better stability, and the composition of their cathode was initially believed to range reversibly between $x$ = 0.067 and 1.0 in $Li_xCoO_2$ within a single non-stoichiometric phase. This classical material, and its derivatives discussed below, subsequently became the cathode of choice for future lithium-ion batteries.

One of the attractions of $Li_xCoO_2$ is that the lithium and cobalt cations are ordered within alternating layers of the crystal structure [**Figure 1(d)**], which facilitates relatively short Co-Co distances that enable good electronic conductivity, an important requirement for battery electrodes. In fact, it has been shown that $Li_xCoO_2$ becomes metallic on charging.[56] It also has a high operating voltage of ≈4 V due to the highly oxidizing nature of $Co^{4+}$. In addition, because the as-synthesised cathode contains lithium, it can be paired with a graphite anode rather than a lithium metal one. Nevertheless, $Li_xCoO_2$ is not as reversible as the 1980 paper had believed because oxygen gas is evolved when the cathode is repeatedly charged beyond ≈$Li_{0.5}CoO_2$ due to the overlap between the $Co^{3+/4+}$ $d$ levels and the top of the $O^{2-}$ $2p$ band.[57] This reduces its practical capacity to about ≈140 mA h g$^{-1}$. In addition, because cobalt is relatively expensive and its availability from mines in the Democratic Republic of the Congo has been hampered by geopolitical issues, alternative $Li_xMO_2$ ($M$ = 3$d$ transition metal) phases with the same crystal structure as $Li_xCoO_2$ have been intensively studied over the last two decades. However, no other single element can approach the performance of cobalt in $Li_xCoO_2$, so most of the recent focus has been on the complex system $LiNi_{1-y-z}Mn_yCo_zO_2$ (known as NMC), especially on compositions that are low in cobalt. The optimal composition is a balance between five factors: chemical stability, structural stability, electrical conductivity, elemental abundance and environmental friendliness, as discussed in a recent review by Manthiram.[58] The general trend is to reduce the cobalt content of the cathode while increasing the nickel content, thereby lowering the cost while increasing the capacity towards 300 mA h g$^{-1}$.[59]

In addition to the extensive work on the layered transition metal oxides, significant progress is also being made on other classes of cathodes for LIBs, especially those based on spinels and polyanion oxides such as $LiFePO_4$.[58] By contrast, there has been relatively little progress in the area of alternatives to the widely used graphite anodes. Recently, however, $TiNb_2O_7$ [**Figure 1(f)**] has emerged as a new anode for fast-charging LIBs.[60] This interesting material presents another striking example of repurposing a compound that has been known for many decades. In this instance, $TiNb_2O_7$ was first identified at the Batelle Research Institute in the early 1950s for potential applications in the nuclear industry, and the $TiO_2$-$Nb_2O_5$ phase diagram was reported shortly afterwards.[61] However, the potential of $TiNb_2O_7$ for use as an anode in LIBs was not recognised until almost 60 years later.[62]

*The role of high throughput synthesis and data/AI/ML in materials creation*

The history of high-throughput or combinatorial synthesis in materials science began in the 1990s with efforts at companies such as Symyx in California and HTE in Germany, as well as academia. There were some early publications in the area,[63,64] but there has not been a great deal of visible impact in terms of new materials. One notable exception is the discovery of metal-organic frameworks (MOFs) that were optimized for $CO_2$ capture by working within a well-defined family of candidates in 9600 microreactions.[65] More recently the area has been re-energised by its integration with exciting developments in computational materials science, as discussed below.

It is interesting to consider the increasing, dual role that artificial intelligence (AI) and machine learning (ML) combined with data mining play in materials discovery. The first are contributions to materials synthesis. In the world of organic chemistry, the combination of tractable (and transferable) valence rules associated with carbon and the lighter main group elements, coupled with the ability to carry out systematic and stepwise bond breaking and making, has allowed much of organic synthesis to be formalized. Since the 1960s, this has led to the development of computer-aided synthesis[66] including, increasingly, the use of Artificial Intelligence/Machine Learning (AI/ML) methods.[67,68]

Extended inorganic compounds, in contrast, are based on the entirety of the periodic table and usually form thermodynamic products, rather than kinetic products. Consequently predictive, controllable synthesis has evolved more slowly. Recently, the use of AI and ML methods based on full text learning of the literature have emerged. These methods have identified synthetic conditions, usually for known materials,[69] with the potential to extend to unknown materials. An unusual approach called the "Dark Reaction Project", is based on learning from successes (as is typically embodied in the published literature) as well as failures (from prospecting in lab notebooks) to arrive at the conditions that would be most likely to yield synthesis products, albeit within a restricted chemical parameter space.[70] The most common use of prediction has become to employ high-throughput quantum mechanical calculations of formation energies to predict potentially stable compounds that have not been previously reported. Such studies do not usually prescribe a synthesis procedure. One of the challenges of these approaches has been in regard to dealing with metastability, and the fact that many materials require kinetic trapping.[71] Advances in the computational understanding the pathways to metastable materials are indeed being made.[72] In terms of thermodynamic products, an early attempt predicted stable four-element alloys.[73] More recently, computation energetics have been employed in conjunction with ML methods, as exemplified by the prediction of unreported ternary oxides with the general composition $A_xB_yO$.[74] An increasingly popular approach is to focus on specific structural classes of materials, such as the Heusler compounds that are known to display a rich range of properties, to learn from known compounds in order to extrapolate to the as-yet unknown.[75] We see great potential in such approaches, particularly when combined with high-throughput synthesis using mobile robots.[76]

These computational approaches to new materials require large and searchable databases of known compounds; the databases have, in turn, **been populated though informed prospecting in chemical parameter space by generations of researchers**. The origins of structural databases can be traced to the work of Bernal and Kennard, and the founding of the Cambridge Crystallography Data Center,[77] which inspired the Inorganic Crystal Structure Database (ICSD).[78] There is also a dedicated database of experimental MOF structures.[79] When the structure/compositions of such databases are fed into a high-throughput computational framework to obtain electronic structures and materials properties, ensuing databases, exemplified by the Materials Project,[80] have become essential components in aiding the screening of materials for interesting functional properties.

*Outlook*

Over the past 150 years, synthetic chemists have built a repository of compositions and crystal structures that now run into the hundreds of thousands. Materials scientists have access to this vast body of information and can explore it in the search for novel functionalities. At the same time, computational tools enable us to interrogate these compounds automatically and to predict their properties. It is reasonable to ask whether we already have enough compositions and structures; is there a need to synthesise more? We would argue, on the contrary, that it remains essential to continue adding to this chemical repository to give ever more options to those who will search for the next generation of functional materials. There is typically a lag of many decades between the initial synthetic discovery and the subsequent applications (**Figure 2**), and materials scientists in future years will undoubtedly find important functionalities in the materials that are being discovered by today's synthetic chemists.

We offer the following general conclusions:

- Many of the most important materials advances involve compounds that were originally synthesised out of curiosity or for entirely different applications. It is important that chemists should continue to contribute to the chemical repository for the benefit of future generations of materials scientists.
- Digital resources that are readily searchable, such as the Inorganic Crystal Structure and the Cambridge Crystallography Data Centre Databases, will continue to play key roles, both for experimental and computational materials scientists who are seeking new functionalities.
- Data-driven materials discovery using AI/ML techniques will make increasingly important contributions, but their success will be greatly enhanced by expanding the current databases with new compositions of matter and crystal structures.

- Funding agencies need to recognise that curiosity-driven synthetic chemistry will be essential for the discovery of many next-generation functional materials.

Finally, it is interesting to explore whether we can already see the beginnings of new materials applications among the novel compounds that have been reported more recently, or among compounds that have been known for many years but are currently attracting a lot of renewed attention. We do not have a crystal ball, but we do see exciting potential in several areas. In the field of quantum materials, the recently-discovered Kagome structure, $CsV_3Sb_3$,[81] is already showing exciting promise as a $Z_2$ topological metal with a superconducting ground state.[82] In the area of halide perovskite photovoltaics, the compound $(NH_4)_3Bi_2I_9$, which was proposed recently as a materials candidate for lead-free PVs,[83] is reported to have outstanding potential for highly sensitive X-ray detection.[84] And in terms of well-established materials seeing a new lease of life, we note that the NaSICON materials, which were first reported in 1968,[85] are now receiving a lot of attention in relation to the development of solid-state sodium ion batteries.[86]